\documentclass{article}
\usepackage{dcase2022,amsmath,graphicx,url,times,booktabs, tabularx}
\usepackage{amsmath,graphicx}
\usepackage{url, times, booktabs, tabularx, amssymb, bm, mathrsfs, mathtools, bbm}
\usepackage{calrsfs, color}
\usepackage{booktabs}
\usepackage{array}
\usepackage{enumitem}
\usepackage{soul}
\usepackage{cite}
\usepackage{siunitx}
\usepackage{adjustbox}
\usepackage{hhline}
\usepackage{makecell}
\usepackage{caption}
\usepackage{float}
\usepackage{stfloats}
\usepackage{tablefootnote}
\usepackage{scrextend}
\usepackage{multirow}
\title{sound event localization and detection for real spatial sound scenes: event-independent network and data augmentation chains}

\name{Jinbo Hu$^{1,2}$, Yin Cao$^{3}$, Ming Wu$^{1}$, Qiuqiang Kong$^{4}$, Feiran Yang$^{1}$, Mark D. Plumbley$^{5}$, Jun Yang$^{1,2}$}
\address{ $^{1}$Key Laboratory of Noise and Vibration Research, Institute of Acoustics, \\
Chinese Academy of Sciences, Beijing, China, \{hujinbo, mingwu, feiran, jyang\}@mail.ioa.ac.cn\\
$^{2}$University of Chinese Academy of Sciences, Beijing, China\\
$^{3}$Xi'an Jiaotong Liverpool University, Suzhou, China, yin.k.cao@gmail.com\\
$^{4}$ByteDance Shanghai, China, kongqiuqiang@bytedance.com\\
$^{5}$Centre for Vision, Speech and Signal Processing (CVSSP), University of Surrey, UK \\
m.plumbley@surrey.ac.uk
}

\begin{document}

\ninept
\maketitle

\begin{sloppy}

\begin{abstract}
Sound event localization and detection (SELD) is a joint task of sound event detection and direction-of-arrival estimation. In DCASE 2022 Task 3, types of data transform from computationally generated spatial recordings to recordings of real-sound scenes. Our system submitted to the DCASE 2022 Task 3 is based on our previous proposed Event-Independent Network V2 (EINV2) with a novel data augmentation method. Our method employs EINV2 with a track-wise output format, permutation-invariant training, and a soft parameter-sharing strategy, to detect different sound events of the same class but in different locations. The Conformer structure is used for extending EINV2 to learn local and global features. A data augmentation method, which contains several data augmentation chains composed of stochastic combinations of several different data augmentation operations, is utilized to generalize the model. To mitigate the lack of real-scene recordings in the development dataset and the presence of sound events being unbalanced, we exploit FSD50K, AudioSet, and TAU Spatial Room Impulse Response Database (TAU-SRIR DB) to generate simulated datasets for training. We present results on the validation set of Sony-TAu Realistic Spatial Soundscapes 2022 (STARSS22) in detail. Experimental results indicate that the ability to generalize to different environments and unbalanced performance among different classes are two main challenges. We evaluate our proposed method in Task 3 of the DCASE 2022 challenge and obtain the second rank in the teams ranking. Source code is released\footnote{\url{https://github.com/Jinbo-Hu/DCASE2022-TASK3}}. 
\end{abstract}


\section{Introduction}
\label{sec:intro}

Sound event localization and detection (SELD) consists of sound event detection (SED) and direction-of-arrival (DoA) estimation. SED aims to detect the presence and types of sound events, and DoA estimation predicts the spatial locations of different sound sources. SELD characterizes sound sources in a spatial-temporal manner. SELD plays an important role in a wide range of applications, such as robot auditory and surveillance of intelligent home.

SELD has received broad attention recently. Adavanne et al.\cite{Adavanne2018_JSTSP} proposed a polyphonic SELD approach using an end-to-end network, SELDnet, which was utilized for a joint task of SED and regression-based DoA estimation. SELD was then introduced in Task 3 of the Detection and Classification of Acoustics Scenes and Events (DCASE) 2019 Challenge for the first time, which uses the TAU Spatial Sound Events 2019 dataset\cite{dcase2019task3}. Most datasets of spatial sound events are computationally simulated and these recordings are generated by convolving randomly chosen sound event examples with a corresponding random real-life spatial room impulse response (SRIR) to spatially place them at a given position\cite{dcase2019task3,dcase2020task3,dcase2021task3}. To bring each iteration of Task 3 of DCASE Challenge closer to real conditions, stronger reverberation, diversity of environment, dynamic scenes with both moving and static sound sources, ambient noise, sound events of the same type, and unknown directional interfering events out of the target classes were added into datasets to complicate the SELD task. In 2022, the challenge transforms from computationally simulated spatial recordings to real spatial sound scene recordings. The Sony-TAu Realistic Spatial Soundscapes 2022 (STARSS22) dataset is manually annotated and released to serve as the development and evaluation dataset of DCASE2022 Task 3 this year\cite{starss22}.

SELDnet is unable to detect sound events of the same type but with different locations\cite{Adavanne2018_JSTSP}, which is also called homogeneous overlap. An event-independent network (EIN) with a track-wise output format was proposed to detect the homogeneous overlap problem \cite{cao2020event,cao2021,hu2022track}. In EIN, there are several event-independent tracks, and each track can be of any event. The number of tracks needs to be pre-determined according to the maximum number of overlapping events. EINV2, an improved version of EIN, utilizes multi-head self-attention (MHSA) and a soft parameter-sharing strategy of multi-task learning to achieve better performance\cite{cao2021}.

The training set often deviates from real-scene spatial and acoustical environments, and mismatched distribution of locations and sound types between the training set and test set are common. A novel data augmentation method is used to generalize the model\cite{hendrycks2019augmix,hu2022track}. The data augmentation method contains several data augmentation chains. These data augmentation chains consist of some randomly sampled data augmentation operations. The augmentation method can increase the diversity of augmented features.

In this study, our system is based on our previous proposed EINV2 with data augmentation chains. EINV2 is extended by Conformer, which is a combination structure of self-attention and convolution. The data augmentation method is composed of several augmentation operations. These data augmentation operations are sampled and layered randomly to combine to several data augmentation chains\cite{hu2022track}. External data is allowed in this challenge. We generate simulated data by stochastically convolving chosen samples of sound events from AudioSet\cite{gemmeke2017audio} and FSD50K\cite{fonseca2021fsd50k} with measured SRIRs from TAU Spatial Room Impulse Responses Database\footnote{\url{https://doi.org/10.5281/zenodo.6408611}} (TAU-SRIR DB). The experimental results show the proposed model with the novel data augmentation method, which was trained on our simulated data, outperforms the DCASE2022 challenge Task 3 baseline model which was trained on official synthetic SELD mixtures\footnote{\url{https://doi.org/10.5281/zenodo.6406873}}. In addition, we present class-wise and room-wise metric scores of the validation set of STARSS22 in detail. The proposed system obtains the second rank in Task 3 of DCASE 2022 Challenge\footnote{\url{https://dcase.community/challenge2022}}.

\section{The method}
\label{sec:method}

\subsection{Input features}
In this method, log-mel spectrograms and intensity vectors (IV) in log-mel space are used for features of the SELD task. First order ambisonics (FOA) include four-channel signals, i.e., omni-directional channel $\mathbf{w}$, and three directional channels $\mathbf{x}$, $\mathbf{y}$, and $\mathbf{z}$. Log-mel spectrograms are computed from the mel filter banks and the short-time Fourier transform spectrograms, and IVs are cross-correlation of log-mel spectrograms of $\mathbf{w}$ with $\mathbf{x}$, $\mathbf{y}$ and $\mathbf{z}$\cite{grumiaux2022survey}. These features are directly calculated online using a 1-D convolutional layer, which supports data augmentation on raw waveform.

\subsection{Network Architecture}

The track-wise output format was introduced in our previous works\cite{cao2020event,cao2021,hu2022track}. It can be defined as
\begin{equation}
\resizebox{0.9\hsize}{!}{$\boldsymbol{Y}_{\text {Trackwise}}=\left\{\left(y_{\mathrm{SED}}, y_{\mathrm{DoA}}\right) \mid y_{\mathrm{SED}} \in \mathbb{O}_{\mathbf{S}}^{M \times K}, y_{\mathrm{DoA}} \in \mathbb{R}^{M \times 3}\right\}$}
\end{equation}
where $M$ is the number of tracks, $K$ is the number of sound-event types, $\mathbb{O}_{\mathbf{S}}^{M \times K}$ is one-hot encoding of $K$ classes, and $\mathbf{S}$ is the set of sound events. Cartesian DoA estimation is used here.

The number of tracks is determined by the maximum polyphony. Each track can only detect a sound event with a corresponding direction of arrival. While a model with a track-wise output format is trained, sound events may be predicted in any track, instead of a fixed track. It may cause the track permutation problem that sound events predicted and their ground truth may not be aligned in a fixed track. Permutation-invariant training (PIT) is proposed to tackle the problem effectively. The PIT loss is defined as

\begin{equation}
\mathcal{L}_{P I T}(t)=
\min _{\alpha \in \mathbf{P}(t)} \sum_{M}\left\{\lambda\cdot\ell^{\mathrm{SED}}_{\alpha}(t)+(1-\lambda)\cdot \ell^{\mathrm{DoA}}_{\alpha}(t)\right\}
\end{equation}
where $\alpha\in \mathbf{P(t)}$ indicates one of the possible permutations and $\lambda$ is a loss weight between SED and DoA. $\ell^\mathrm{SED}_\alpha$ is binary cross entropy loss for the SED task, and $\ell^{\mathrm{DoA}}_\alpha$ is mean square error for the DoA task. The lowest loss will be chosen by finding a possible permutation, and the back-propagation is then performed. 

From multi-task learning (MTL) perspective, joint SELD learning can be mutually beneficial. Hard parameter-sharing (PS) and soft PS are two typical methods to implement MTL. Hard PS means subtasks use the same feature layers, while soft PS means subtasks use their own feature layers with connections existing among those feature layers. In \cite{cao2021}, experimental results show that soft PS using cross-stitch is more effective.

EINV2, which combines the track-wise output format, PIT, and soft PS, is utilized in our system. Three tracks are adopted to address up to three overlapped sound events. Multi-head self-attention (MHSA) blocks are replaced with Conformer blocks. Conformer consists of two feed-forward layers with residual connections sandwiching the MHSA and convolution modules, and hence has the ability to capture global and local patterns.\cite{gulati2020conformer,hu2022track}. Our proposed network is shown in Fig. \ref{fig: single model}.

\begin{figure}[tb]
  \centering
  \scalebox{1.0}{\centerline{\includegraphics[width=\columnwidth]{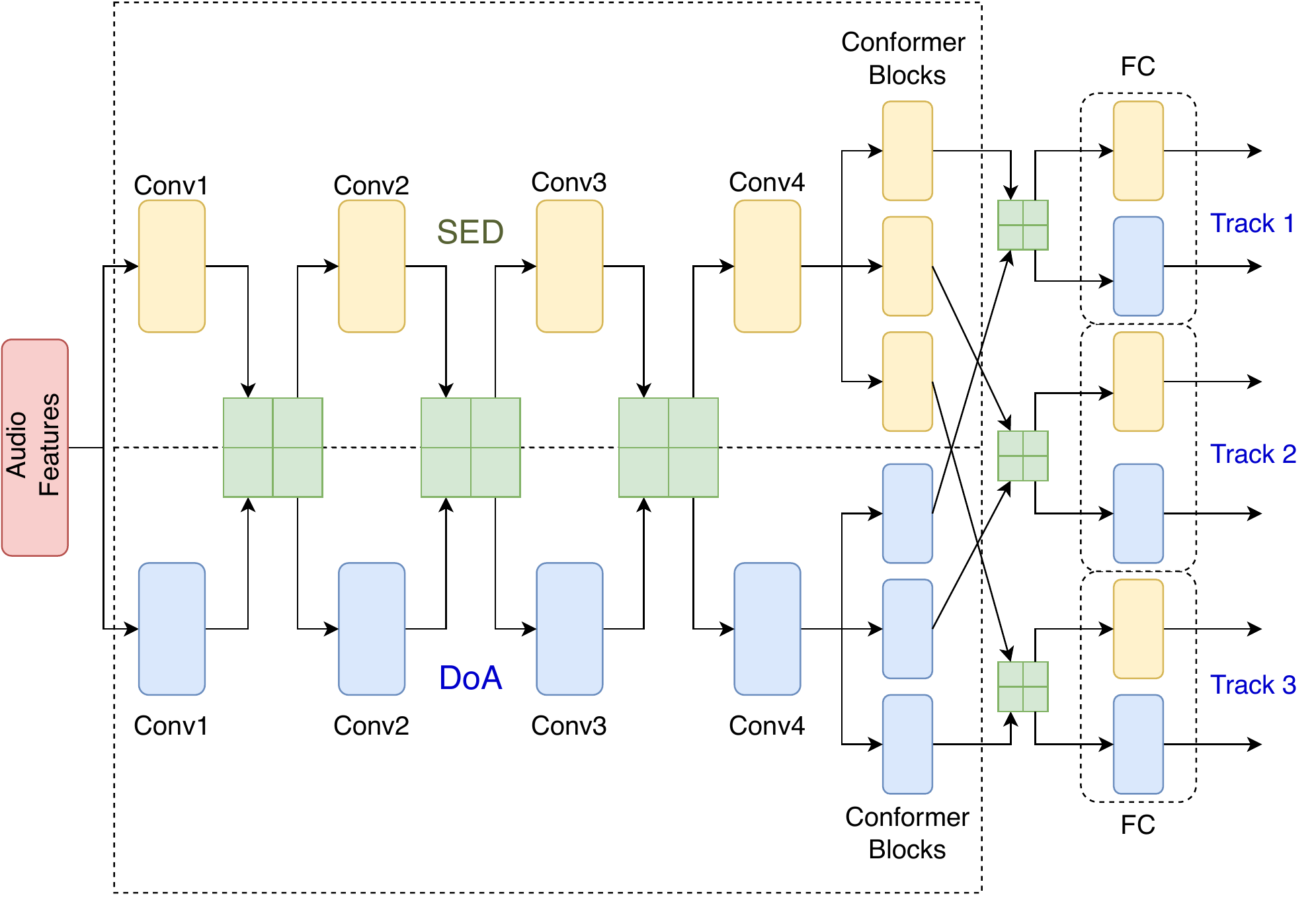}}}
  \vspace{-5mm}
  \caption{The architecture of the SELD network, which is a Conv-Conformer network. The upper half (yellow boxes) is the SED task. The lower half (blue boxes) is the DoA estimation task. The green boxes sandwiched between SED branch and DoA branch indicate soft connections between SED and DoA estimation.}
  \label{fig: single model}
  \vspace{-5mm}
\end{figure}

\subsection{Data Augmentation Chains}
The main characteristic of our data augmentation method is using some augmentation chains\cite{hendrycks2019augmix, chen2020simple, hu2022track}. These augmentation chains are combined by some augmentation operations, which are randomly selected and linked in chain. We randomly sample $k=3$ augmentation chains. Augmentation operations that are used here include Mixup\cite{zhang2018mixup}, Cutout\cite{zhong2020random}, SpecAugment\cite{specaug}, and frequency shifting\cite{nguyen2021salsa}. Rotation of FOA signals\cite{mazzon2019first} is an additional augmentation method, but excluded by data augmentation chains. 
The diagram of data augmentation chains is shown in Fig. \ref{fig: dataAug chains}.

Mixup utilize convex combinations of pairs of feature vectors and their labels to train the model. Mixup on both raw waveform and spectrograms is used here to improve the ability of detecting overlapping sound events. While random Cutout produces several rectangular masks on spectrograms, SpecAugment produces stripes masks on time and frequency dimension of spectrograms. Frequency shifting in the frequency domain is similar to pitch shift in the time domain, and it randomly shifts input features of all the channels up or down along the frequency dimension by several bands. We also use a spatial augmentation method, rotation of FOA signals. It rotates FOA format signals by channel swap to enrich DoA labels. This method does not lose physical relationships between sound sources and observers. We use z-axis as the rotation axis to swap directional channel $\mathbf{x}$ and $\mathbf{y}$, which leads to 16 types of channel rotation.

\begin{figure}[tb]
  \centering
  \vspace{-3mm}
  \scalebox{1.0}{\centerline{\includegraphics[width=\columnwidth]{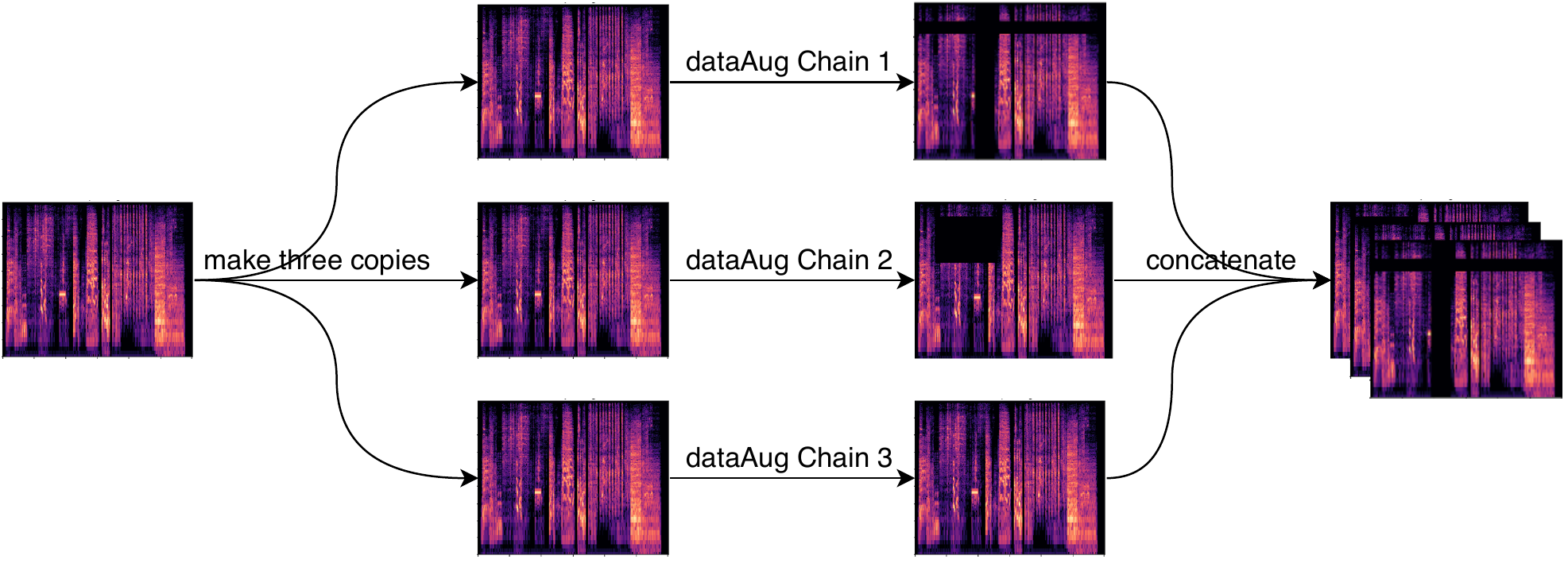}}}
  \caption{Diagram of data augmentation chains}
  \label{fig: dataAug chains}
  \vspace{-7mm}
\end{figure}

\subsection{Simulated Data}
Manual annotations are expensive and the duration of STARSS22 (about 5 hours of the development set) is limited compared with the synthetic datasets (about 13 hours synthetic recordings in DCASE 2021) used in previous years, therefore, external datasets are used to improve the model performance. We generated simulated data using the generator code\footnote{\url{https://github.com/danielkrause/DCASE2022-data-generator}} provided by DCASE 2022. 

Samples of sound events are mainly sourced from FSD50K dataset, based on affinity of the labels in that dataset to the target classes. The target class \textit{background music} and the interference class \textit{shuffling cards} are not in FSD50K dataset, therefore, we use AudioSet as a supplement. Spatial events were spatialized in 9 unique rooms, using collected SRIRs from the TAU-SRIR DB dataset. The ambient noise from the same room was additionally mixed at varying signal-to-noise ratios (SNR) ranging from 30 dB to 6 dB. The maximum polyphony of target classes is 3, excluding additional polyphony of interference classes. 

We select sound event samples whose labels significantly corresponded to the target classes. Each sound event sample also has a different energy gain for mixing. By setting different ranges of gain and choosing different samples, we generate three datasets, A, B, and C. All of these synthetic datasets have 2700 1-minute clips.

\section{Experiments}
\begin{table*}[ht]
    \centering
    \caption{The SELD performance of our proposed system. The training set of STARSS22 is mixed into synthetic training set by default.}
    \label{tab:results}
    \resizebox{\textwidth}{!}{
        \begin{tabular}{c|c|cccc|cccc}
        \toprule
        &  & \multicolumn{4}{c|}{Validation set} & \multicolumn{4}{c}{Evaluation (Blind test) set} \\
        System & Datasets & $\mathrm{ER}_{20^{\circ}}$ & $\mathrm{F}_{20^{\circ}}$ & $\mathrm{LE}_\mathrm{CD}$ & $\mathrm{LR}_\mathrm{CD}$ & $\mathrm{ER}_{20^{\circ}}$ & $\mathrm{F}_{20^{\circ}}$ & $\mathrm{LE}_\mathrm{CD}$ & $\mathrm{LR}_\mathrm{CD}$ \\
        \midrule
        Baseline FOA~\cite{starss22} & Official & 0.71 & 21.0\% & $29.3^{\circ}$ & 46.0\% & 0.61 & 23.7\% & $22.9^{\circ}$ & 51.4\%   \\
        EINV2 w/o dataAug chains& Official & 0.75 & 32.3\% & $24.0^{\circ}$ & 56.1\% & - & - & - & - \\
        EINV2 w/\quad dataAug chains & Official & 0.56 & 42.4\% & $19.3^{\circ}$ & 61.4\% & - & - & - & - \\  % E1.1 OS
        \midrule
        System \#1 & A+B+C & 0.50 & 48.4\% & $19.5^{\circ}$ & 65.7\% & 0.44 & 49.2\% & $16.6^{\circ}$ & 70.4\%  \\  
        System \#2 & A+B & 0.50 & 51.0\% & $16.4^{\circ}$ & 65.9\% & 0.40 & 57.4\% & $15.1^{\circ}$ & 70.6\%  \\  
        System \#3 & A & 0.53 & 48.1\% & $17.8^{\circ}$ & 62.6\% & 0.39 & 55.8\% & $16.2^{\circ}$ & 72.4\%  \\  
        System \#4 & B & 0.53 & 45.4\% & $17.4^{\circ}$ & 62.5\% & 0.40 & 50.9\% & $15.9^{\circ}$ & 69.4\%  \\  
        \bottomrule
        \end{tabular}
        }
    \vspace{-3mm}
\end{table*}

\subsection{Datasets}
The STARSS22 dataset contains recordings of real scenes, and the density of sound event samples and the presence of each class varies greatly. The maximum number of the overlaps is 5, but those cases are very rare\cite{starss22}. The overlap of 4 and 5 accounts for the proportion of 1.8\% in total. Occurrences of up to 3 simultaneous events are fairly common, so we ignore the case scenarios that the number of overlapping events is more than 3. During the development stage, we train our proposed model on mixed datasets of synthetic recordings and the training set of STARSS22, and evaluate those systems using the validation set of STARSS22. During the evaluation stage, both synthetic recordings and all of the development set of STARSS22 are used for training.

\subsection{Hyper-parameters}
Audio clips are segmented to have a fixed length of 5 seconds with no overlap for training and inference. Log-mel spectrograms and intensity vectors features, with 24 kHz sampling rate, a 1024-point Hanning window with a hop size of 400, and 128 mel bins, are extracted from these audio segments. AdamW optimizer is used. The learning rate is set to 0.0003 for the first 70 epochs and is decreased to 0.00003 for the following 20 epochs. The threshold for SED is set to 0.5 to binarize predictions. The loss weight $\lambda$ is 0.5.

\subsection{Evaluation Metrics}
We use the official evaluation metrics to evaluate the SELD performance\cite{overviewofDCASE,mesaros2019joint}. The evaluation metrics use a joint metric of localization and detection: location-sensitive F-score ($\mathrm{F}_{\leq T^\circ}$), error rate ($\mathrm{ER}_{\leq T^\circ}$), and class-sensitive localization recall ($\mathrm{LR}_\mathrm{CD}$), localization error ($\mathrm{LE}_\mathrm{CD}$). $T^\circ$ means spatial threshold and is set to $20^\circ$ in this challenge. $\mathrm{F}_{\leq T^\circ}$ and $\mathrm{ER}_{\leq T^\circ}$ consider true positives predicted under a spatial threshold $T^\circ$  from the ground truth. For $\mathrm{LE}_\mathrm{CD}$ and $\mathrm{LR}_\mathrm{CD}$, the detected sound class has to be correct in order to count the corresponding localization predictions. 

Contrary to the previous challenges, where the evaluation metrics were micro-averaged, this gives equal weight to each individual decision and the performance is affected by the classes with more samples. In this challenge, macro-averaging of evaluation metrics is used. Macro-averaging gives equal weight to each class and emphasizes the system performance on the smaller classes\cite{SEDMetric}.

\subsection{Experimental Results}

\begin{figure*}[h]
  \centering
%   \vspace{-2mm}
  \scalebox{1.0}{\centerline{\includegraphics[width=\textwidth]{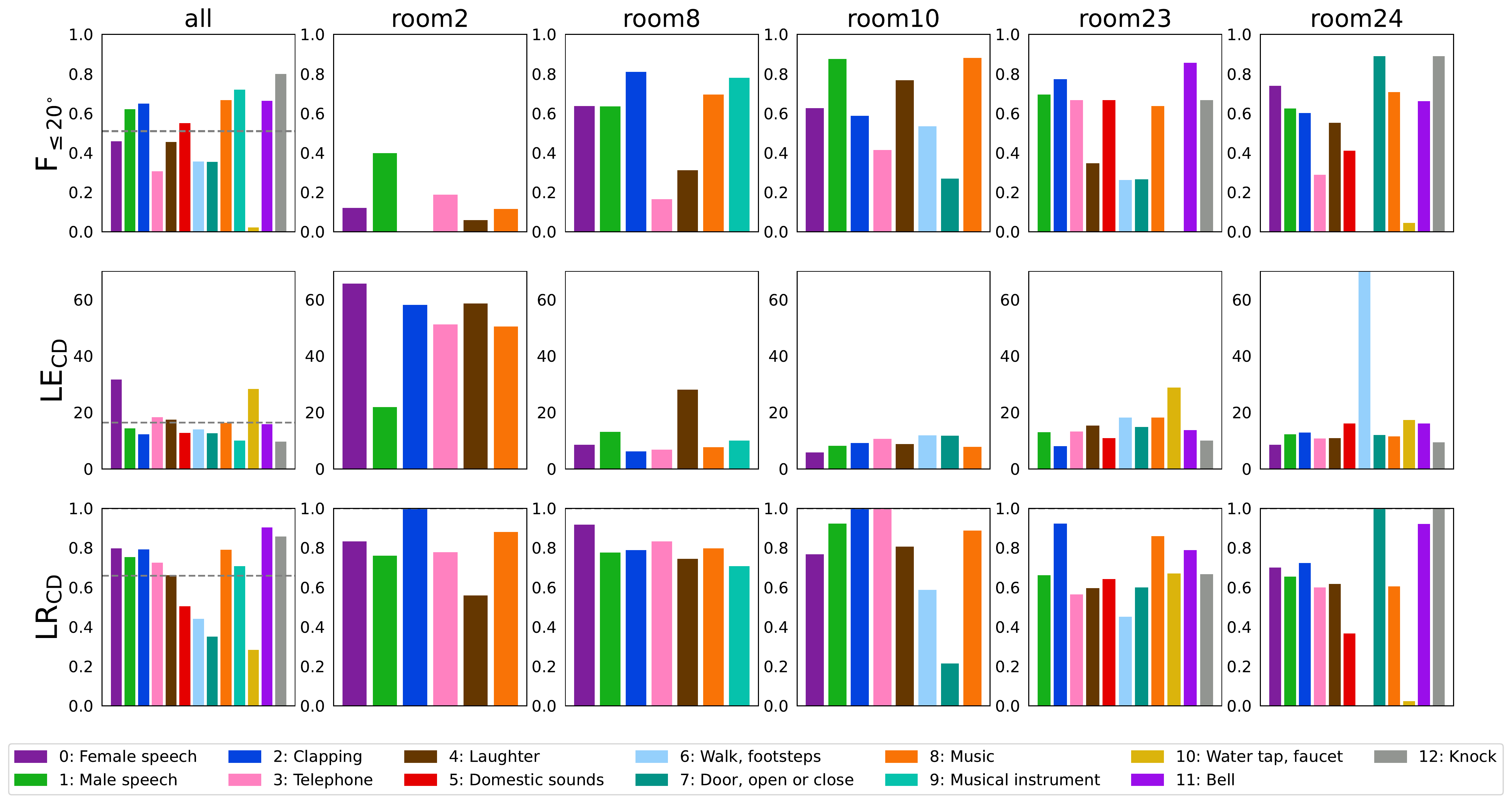}}}
  \caption{Metric scores of System \#2 on validation set of STARSS22 in detail. The first column shows metric scores of the whole validation set. The following columns present metric scores of each room of validation set.}
  \label{fig: score_aggregation}
%   \vspace*{-5mm}
\end{figure*}

Table \ref{tab:results} summarizes the performance of our proposed systems. The official dataset means the synthetic mixtures for baseline training. The system baseline, EINV2 without dataAug chains, and EINV2 with dataAug chains all use the same dataset for training. EINV2 without data augmentation chains outperforms the baseline model, whereas EINV2 with data augmentation chains performs better. 

All configurations of systems \#1 - \#4 are the same as system EINV2 with dataAug chains, except for the training set used. The results also demonstrate the effectiveness of our simulated data over the official dataset.

The first column of Fig. \ref{fig: score_aggregation} shows class-wise metric scores of System \#2 on the validation set of STARSS22. The class-wise performance on the whole validation set is highly skewed, with $\mathrm{F}_{\leq 20^\circ}$ of \textit{knock} class being 80.0\%, whereas $\mathrm{F}_{\leq 20^\circ}$ of \textit{water tap and faucet} class being 2.2\%. $\mathrm{LE}_\mathrm{CD}$ of \textit{female speech} class and \textit{water tap and faucet} class is a lot higher than average. Other columns of Fig. \ref{fig: score_aggregation} present class-wise performance for each room. Unbalanced class-wise performance among different rooms results in the skewed class-wise performance on the whole validation set.

The performance of the localization in room 2 is the worst among all the rooms, resulting in a directly significant increase of $\mathrm{LE}_\mathrm{CD}$ of \textit{female speech} class. It may be attributed to small room size of room 2 compared with other rooms. $\mathrm{LR}_\mathrm{CD}$ of \textit{walk, footsteps} (0.0\%) class and \textit{water tap and faucet} (2.4\%) class in room 24 is very low. A possible reason is the low quality of synthetic training samples, because we ignore the natural temporal occurrences and spatial connections of some types of sounds happening in real scenes when simulating data\cite{starss22}. For example, the target class \textit{water tap and faucet} and the directional interference class \textit{dishes, pots, and pans} often occur simultaneously in room 24, which leads to many observed false negatives of the class \textit{water tap and faucet} in the system output. It is difficult to synthesis training samples that contains the temporal and spatial relationships of sound events in real scenes. These factors can lead to performance degradation.

\section{Conclusion}

We have presented an approach using an Event-Independent Network V2 (EINV2) with a novel data augmentation method for real-life sound event localization and detection. EINV2 is extended by conformer blocks. The novel data augmentation method contains several augmentation chains, which are stochastic combinations of data augmentation operations. For this challenge, we synthesized more training samples which are convolved using sound events from FSD50k and AudioSet with measured room impulse responses from TAU-SRIR DB. Our model with data augmentation chains performs better than the baseline model. Furthermore, experimental results show further improvement with our synthetic datasets. We also show results on the validation set of STARSS22 in detail. Our proposed method is evaluated in the evaluation set of STARSS22, and obtained the second best team in Task 3 of DCASE 2022 Challenge. The study of the generalization ability to different environments and the performance for unbalanced classes will be analyzed further in the future work.

\section{Acknowledgement}
This work was partly supported by Frontier Exploration project independently deployed by Institute of Acoustics, Chinese Academy of Sciences (No. QYTS202009), UK Engineering and Physical Sciences Research Council (EPSRC) grant EP/T019751/1 “AI for Sound”. For the purpose of open access, the authors have applied a Creative Commons Attribution (CC BY) licence to any Author Accepted Manuscript version arising.

% -------------------------------------------------------------------------
% Either list references using the bibliography style file IEEEtran.bst
\bibliographystyle{IEEEtran}
\bibliography{refs}

\end{sloppy}
\end{document}